\documentclass[notitlepage]{revtex4-1}

\usepackage{amsmath}
\usepackage{amsfonts}
\usepackage{amsthm}
\usepackage{amssymb}
\usepackage{graphicx}
\usepackage{epsfig}
\usepackage{hyperref}
\usepackage{color}

\begin{document}
\title{Relativistic  disks with two charged perfect fluids components}

\author{Antonio C. Guti\'errez-Pi\~{n}eres}
\email[e-mail:]{acgutierrez@correo.nucleares.unam.mx}
\affiliation{Instituto de Ciencias Nucleares, Universidad Nacional Aut\'onoma de M\'exico,
 \\AP 70543,  M\'exico, DF 04510, M\'exico}
\affiliation{Facultad de Ciencias B\'asicas,\\
Universidad Tecnol\'ogica de Bol\'ivar, Cartagena 13001, Colombia}
		
\begin{abstract}
A method to describe exact solutions of the Einstein-Maxwell field equations in terms of relativistic thin disks constituted by two perfect charged fluids is presented. Describing the surface of the disk as a single charged fluid we find explicit expressions for the rest energies, the  pressures and the electric charge densities of the two fluids. An explicit example is given. The  particular case of the thin disks composed by  two charged perfect fluids  with barotropic equation of  state is also presented.
\end{abstract}

\maketitle
\section{\label{sec:intro}Introduction}
The  problem of constructing exact  models in general relativity to describe axially symmetric sources from the astrophysical point of view remains  a topical problem. In particular, finding exact solutions of the Einstein-Maxwell field equations is a difficult task due to the non-linear character of the equations, in fact there are only a handful of  physically acceptable  solutions. 
In order to construct the  models  and  obtain the  solutions, one can follow two approaches. On the one hand, an equation of state is given and then the field equations are solved. On the other hand by means of the inverse method, where a solution of the field equations is taken and then the energy-momentum (EM) tensor is obtained \cite{GG-P2012}. For both cases, one  has to pose correctly and solve for the exterior and the interior given boundary conditions for the Einstein-Maxwell equations. To obtain  physically reasonable models the solutions must satisfy additional conditions compatible with the observable nature.

The energy-momentum tensor obtained using the inverse method has not a direct physical interpretation. Hence, to understand the physical meaning of the EM tensor, the  eigenvalue problem should be solved. Alternatively, a suitable tetrad can be used to establish a relationship between the physical features of the tensor obtained from the inverse method and those of a well-known energy-momentum tensor. In most of the cases, the physical interpretation is difficult to accomplish. The inverse method has led to solutions of the Einstein-Maxwell equations in terms of thin disk models whose energy-momentum tensor can be  interpreted as that of a single charged dust fluid, see for example \cite{GG-PO,G-PGQ} and  references therein. 
The use of two fluids models to describe the surface energy-momentum tensor of relativistic disks and the understanding of the  effects caused by the motions of each fluid component, can  provide insight into the description of the sources for astrophysical objects. Two component anisotropic fluids have been studied in \cite{Letelier1980, Krisch-Glass} and with multi-fluid components in \cite{Letelier1986}. The problem of the relativistic thin disks constituted by only two non-interacting fluids has been explored in \cite{G-PL-MQ}.

Inspired by the work of Letelier \cite{Letelier1980}, we present a simple method to interpret the material source of the relativistic thin disks in terms of two charged-perfect-fluids. As a direct application of the method, we study the solution to the Einstein-Maxwell equations for the thin disk-halos obtained in \cite{G-PL-MQ}. This paper is organized as follows. First, in section \ref{sec:model} we introduce the two charged perfect fluid components method, where the  surface energy-momentum tensor and the surface electric current density of the relativistic thin disk can be interpreted as a two perfect charged fluids. In the following section \ref{sec:D-HCPFC} we apply the method here presented to the solutions of the Einstein-Maxwell equations corresponding to thin disk-halos systems, as those in \cite{G-PL-MQ}, to describe relativistic thin disks in terms of two charged perfect fluids. Finally, we present some concluding remarks in section \ref{sec:conclus}.

\section{\label{sec:model} The two charged perfect fluid components  method}
In this section, we present a method for interpreting the solutions of the Einstein-Maxwell field equations corresponding to  disk-like configurations of matter in terms of two charged perfect fluid components. To do so, we first assume that the surface energy-momentum tensor (SEMT) and the surface electric current density (SECD) of the disk can be written in the following form
\begin{subequations}\begin{eqnarray}
S^{ab}(u,v) &=& t^{ab}(u) + t^{ab}(v),\label{eq:twoSEMT},\\
J^{a}(u,v) &=& J^{a}(u) + J^{a}(v),\label{eq:twoCURRENT}
\end{eqnarray}\label{eq:twoSEMT-CURRENT} \end{subequations}
where the  tensors  $t^{ab}(u)$ and $t^{ab}(v)$ and the 4-vectors $J^a(u)$ and $J^a(v)$ are defined as
\begin{subequations}\begin{eqnarray}
 t^{ab}(u)&=& (p + w)u^au^b + ph^{ab}\label{eq:SEMT1},\\
 t^{ab}(v)&=& (q + e)v^av^b + qh^{ab} \label{eq:SEMT2},\\
 J^{a}(u)&=& \eta u^a \label{eq:CURRENT1},\\
 J^{a}(v)&=& \lambda v^a \label{eq:CURRENT2}.
\end{eqnarray}\label{eq:SEM12-CURRENT12}\end{subequations}
The SEMT of the disk (\ref{eq:twoSEMT}) is the sum of the surface energy-momentum tensor of two perfect fluids with pressures $p$ and $q$ and rest energies $w$ and $e$, respectively; $h^{ab}$ is  the induced  metric on the surface of the disk. Similarly, the SECD of the disk (\ref{eq:twoCURRENT}) is the sum of two surface electric charge densities $\eta$ and $\lambda$  flowing with 4-velocities $u$ and $v$ on the surface of the disk. The 4-velocities of the fluids satisfy the condition
\begin{eqnarray}
u^au_a = v^av_a &=& -1, \qquad u^a \neq v^a,
\end{eqnarray}
that is, the two time-like  fluid velocities are not necessarily aligned. This  can  lead  to interesting transport effects \cite{L-M1} and to the generation of dynamical instabilities \cite{L-M2}. Now, we analyze the physical meaning of the SEMT and the SECD (\ref{eq:twoSEMT-CURRENT}). 
To this end, we introduce the  transformation
\begin{subequations}\begin{eqnarray}
u^a \longrightarrow  u^{* a} &=& u^a \cos{\alpha}  + v^a \left(\frac{q + e}{p + w}\right)^{1/2} \sin{\alpha} ,\\ 
v^a \longrightarrow  v^{* a} &=&  -u^a \left(\frac{p + w}{q + e}\right)^{1/2} \sin{\alpha}  + v^a \cos{\alpha} ,
\end{eqnarray}\label{eq:transf}\end{subequations}
in order to cast them in the general form of the SEMT and the SECD of a single  fluid \cite{SYN}. 
It is easy to see that the SEMT is invariant under this transformation, then we can write this as 
\begin{eqnarray}
 S^{ab}&=& (p + w)u^{a*}u^{*b} + (q + e)v^{*a}v^{*b} +(p+q)h^{ab}.
 \label{eq:SEM12*}\end{eqnarray}
whereas under this transformation the surface electric current density is
\begin{eqnarray}
J^a = q_1u^{*a} + q_2v^{*a}.\label{eq:current*}
\end{eqnarray}
Here $q_1$ and $q_2$ are the surface electric charge densities of the fluids in the ``new frame''. The rotation angle $\alpha$ in (\ref{eq:transf}) is fixed in terms of the velocity overlap by requiring  $u^{*a}v^*_{\;a} =0$, therefore, we may choose $u^{a*}$ as a time-like vector and $v^{*a}$ as a space-like vector. Equation (\ref{eq:transf}) thus leads  to
\begin{eqnarray}
\tan{2\alpha}=  \frac{(p + w)^{1/2}(q + e)^{1/2}(2u^av_a)}{(q+ e) - (p +w)}
\label{eq:tan}.
\end{eqnarray}
We  are  now  able to write down the  (SEMT) and the (SECD) (\ref{eq:twoSEMT-CURRENT}) in the general form of the energy-momentum tensor and electric current density for a {\it single  fluid} flowing with 4-velocity $U^a$, 
\begin{subequations}
\begin{eqnarray}
S^{ab} &=& \rho U^aU^b - {\tau}^{ab},\label{eq:semtSYNGE1}\\
J^{a} &=& \psi U^a,\label{eq:secdSYNGE2}
\end{eqnarray}\label{eq:semt-secdSYNGE}
\end{subequations}
with
\begin{subequations}
\begin{eqnarray}
\tau^{ab}U_a&=&0,\\
U^aU_a&=&-1.
\end{eqnarray}
\end{subequations} 
where $\rho$ is  the rest energy of the single fluid, $\tau^{ab}$ is  called the {\it stress tensor} and $\psi$ is the surface electric charge density of the fluid. If we define the following unit vectors
\begin{subequations}\begin{eqnarray}
U^a &\equiv & \frac{u^{*a}}{(- u^{*a}u^{*}_{\;a})^{1/2}},\label{eq:U}\\
\chi^a &\equiv & \frac{v^{*a}}{(v^{*a}v^{*}_{\;a})^{1/2}},
\end{eqnarray}\label{eq:Uchivectors}\end{subequations}
which satisfy the conditions
$ U^aU_a = -\chi^a\chi_a=-1$ and $U^a\chi_a=0$, it is  possible to define the  quantities
\begin{subequations}\begin{eqnarray}
\rho &\equiv &  S^{ab}U_aU_b= -(p+w)u^{*a}u^*_{\;a}- (p + q),\\
\gamma &\equiv &  S^{ab}\chi_a\chi_b  = + (q+e)v^{*a}v^*_{\;a} + (p + q),\\
\psi &\equiv& -U^aJ_a =q_1(-u^{*a}u^*_{a})^{1/2},\label{eq:charge}\\
\pi &\equiv & p + q,
\end{eqnarray}\label{eq:pqw**}\end{subequations}
where we have  used the equations (\ref{eq:semt-secdSYNGE}). Using that $S^{ab}(u,v)= S^{ab}(u^*,v^*)$, the surface energy-momentum tensor can be written as 
\begin{eqnarray}
S^{ab} = (\rho + \pi)U^aU^b + (\gamma - \pi)\chi^a\chi^b + \pi h^{ab}, \label{eq:SEMTUchi}
\end{eqnarray}
which can be expressed as the surface energy momentum tensor of a single fluid (\ref{eq:semtSYNGE1}) with rest energy $\rho$  and with stress tensor
\begin{eqnarray}
\tau^{ab} &=& - (\gamma - \pi)\chi^a\chi^b - \pi(U^aU^b + h^{ab}).
\end{eqnarray}
Note  that
\begin{eqnarray*}
\chi^a U_a&=&0,\\	
\tau^{ab}U_a&=&0,\\
\tau^{ab}\chi _{b}&=& - \gamma\chi^a.
\end{eqnarray*} 
Again, by using  the fact that $S^{ab}(u,v) = S^{ab}(u^*,v^*)$ and the equations (\ref{eq:pqw**}), we  obtain the  quantities  $\rho$ and $\gamma$ in terms of $p, w, q$ and $e$,
\begin{subequations}\begin{eqnarray}
\rho&=&+\frac{1}{2}(w + e - p- q) + \frac{1}{2}\left\{ (p + w -q - e)^2 +
4(p+w)(q+e)(u^av_a)^2\right\}^{1/2},\label{eq:rho}\\
\gamma&=&-\frac{1}{2}(w + e -p - q) + \frac{1}{2}\left\{ (p+w -q -e)^2 +
4(p+w)(q+e)(u^av_a)^2\right\}^{1/2},
\end{eqnarray}\label{eq:rho-gamma}\end{subequations}
where  $w$, $p$, $e$ and $q$ are the quantities defined in (\ref{eq:SEM12-CURRENT12}). To write the SECD in the general form for the electric current density of a {\it single  fluid}, we use the  transformation (\ref{eq:transf}) and  the surface electric current density given by (\ref{eq:current*}). Then we obtain 
\begin{eqnarray}
J^a= u^a \left[ q_1 \cos{\alpha} - q_2\left( \frac{p + w}{q + e}\right)^{1/2}\sin{\alpha}\right]
 + v^a \left[ q_1 \left(\frac{q + e}{p + w}\right)^{1/2}\sin{\alpha} + q_2\cos{\alpha} \right] \label{eq:current**}.
\end{eqnarray}
Consequently, from the  equations (\ref{eq:twoCURRENT}) and (\ref{eq:current**}) we  have the  following relation between the surface electric charge densities  
\begin{eqnarray}
(q + e)(q_1^2 - \eta^2)+ (p+w)(q_2^2 - \lambda^2)  = 0.\label{eq:betlamrelat}
\end{eqnarray}
On the other  side, by using this relation, (\ref{eq:rho}) and the definition of  surface charge density given by (\ref{eq:charge}) we  have
\begin{eqnarray}
\psi= \left(  -\frac{q_2^2}{q+e}  +  \frac{\eta^2}{p+w} + \frac{\lambda^2}{q+e} \right)^{1/2}\left( \rho - p - q\right)^{1/2}.
\end{eqnarray}
 Now, in order to guarantee the  invariance of the  electric charge density we demand that
\begin{eqnarray}
\psi=(-J^a(U)J_a(U))^{1/2}= (-J^a(u^*,v^*)J_a(u^*,v^*))^{1/2}=(-J^a(u,v)J_a(u,v))^{1/2}.
\end{eqnarray} 
From this it follows that $q_2=0$, and that the surface charge density $\psi$ defined in (\ref{eq:secdSYNGE2}) becomes
\begin{eqnarray}
\psi= \left( \eta^2 + \lambda^2  - 2\eta\lambda u^av_a  \right)^{1/2}.
\end{eqnarray}
We thus  have the  following relations among the surface electric charge densities, the pressures, the rest energies and the velocities of  the  fluids: 
\begin{subequations}\begin{eqnarray}
\psi &=& \left( \frac{\eta^2}{p+w} + \frac{\lambda^2}{q+e} \right)^{1/2}\left( \rho - p - q\right)^{1/2},\\
2u^av_a&=&\frac{\eta (2p + w + q-\rho)}{\lambda(p+w)}+
\frac{\lambda (2q + e + p -\rho)}{ \eta(q+e)}.
\end{eqnarray}\label{eq:scd-velocity}\end{subequations}
In sum, we are equipped with a set of useful relations  from which we  can interpret the sources of the relativistic thin disk in terms of  two charged perfect fluids. As  we know, through the inverse method, we can always  have explicit expressions for  the surface energy momentum tensor and the surface electric current  density of  the relativistic disks. Then,  we can  always  choose a suitable tetrad in terms of which  the equations (\ref{eq:tan}), (\ref{eq:rho-gamma})  and (\ref{eq:scd-velocity})  can be  used to express the rest energy, the  pressures, the velocity  and the  electric charge density  of a single charged fluid characterizing the relativistic disk, in terms of two charged perfect fluids. In the next section we present a simple example to illustrate the application of the method developed in this section.

\section{\label{sec:D-HCPFC}Relativistic Disk-halos with two charged perfect fluids components}
In order to  illustrate the  application of  the   method presented in the previous section to specific thin disks  solutions, we  will consider the solutions of  the  Einstein-Maxwell equations corresponding to thin disk-halos recently presented in \cite{G-PL-MQ}. This kind of  disks are constituted by a  single charged dust fluid surrounded by a halo in presence of  an  electromagnetic  field. The  metric tensor is  given by the conformastatic line element \cite{SYN}
\begin{eqnarray}
 \mathrm ds^2 = - \mathrm e^{2 \phi} \mathrm dt^2 \ + \mathrm e^{-2\beta\phi}
 [r^2{\mathrm d}\varphi^2  + \mathrm dr^2 + \mathrm
 dz^2]. \label{eq:met0} 
\end{eqnarray}
where $\phi$ depends on $r$ and $z$ only.  Here, we  only analyze the material source on the  surface  of  the  disk. First, we will cast the  surface energy momentum tensor and the surface electric current  density of  the disks in terms of  the locally static observer (LSO), that is 
\begin{subequations}\begin{eqnarray}
S^{ab}_D &=& \epsilon e_{(0)}^{\;\;\;a}e_{(0)}^{\;\;\;b} + \wp e_{(1)}^{\;\;\;a}e_{(1)}^{\;\;\;b} + 
\wp e_{(2)}^{\;\;\;a}e_{(2)}^{\;\;\;b},
\label{eq:SEMTdisk1}\\
J^{a}_D &=& -\sigma e_{(0)}^{\;\;\;a},\label{eq:SEMTdisk2}
\end{eqnarray}\label{eq:SEMTdisk}\end{subequations}
where the non-zero components of  the LSO are given by
\begin{subequations}\begin{eqnarray}
 e_{(0)}^{\;\;a} &=& e^{-\phi} \delta_{0}^{a}, \\ 
 e_{(1)}^{\;\;a}&=& r^{-1}e^{\beta\phi} \delta_{1}^{a}, \\ 
 e_{(2)}^{\;\;a}&=& e^{\beta\phi} \delta_{2}^{a},\\ 
 e_{(3)}^{\;\;a}&=& e^{\beta\phi} \delta_{3}^{a},
\end{eqnarray}\end{subequations}
being $\delta$ the usual Kronecker's delta symbol. With respect to the LSO, the components of the surface energy momentum tensor can be interpreted as the rest energy density and the pressures of the disk, given by the expressions 
\begin{subequations}\begin{eqnarray}
	\epsilon(r) 	 &=& 4\beta ma F(r),\\	
\wp (r) 	& = &\aleph \rho(r),\\
F(r) & =& \frac{{(r^2 + a^2)}^{-\frac{2+\beta}{	2 + 2\beta}}}{(1 + \beta ) \left(\sqrt{r^2 + a^2} - m\right)^{\frac{2\beta +1}{1 + \beta}}}. 
\end{eqnarray}\end{subequations}
while the surface charge density of the  disks is 
\begin{eqnarray}
\sigma(r)=\frac{2k_1ma}
{ k\left(\sqrt{r^2 + a^2} - m\right)^{\frac{2\beta}{1 + \beta}}
 (r^2 + a^2)^{\frac{3 +\beta}{2(1 +\beta)}}}
\end{eqnarray}
 with  $k_1$, $a$, $m$ and $k$ arbitrary constants and $\aleph = {(1 -\beta)}/{2\beta}$. 
 The parameter $\beta$  produces a non-zero pressure and  it has the same value in the radial and angular directions. Now we rewrite (\ref{eq:SEMTUchi}) to  get
\begin{eqnarray}
S^{ab}= \rho U^aU^b + (\gamma - \pi)\chi^a\chi^b + \pi( h^{ab} + U^aU^b).\label{eq:SEMTletB}
\end{eqnarray}
Comparing  (\ref{eq:SEMTdisk1}) and (\ref{eq:SEMTletB}), we  find  that the induced  metric on the  surface of  the  disk can be  expressed as
\begin{eqnarray}
h^{ab}&\equiv& g^{ab} - e_{(3)}^{\;\;\;a}e_{(3)}^{\;\;\;b}= - e_{(0)}^{\;\;\;a}e_{(0)}^{\;\;\;b}+ e_{(1)}^{\;\;\;a}e_{(1)}^{\;\;\;b}+ e_{(2)}^{\;\;\;a}e_{(2)}^{\;\;\;b},\label{eq:hab}
\end{eqnarray}
and that the  vectors  $U^a$ and $\chi^a$ given by (\ref{eq:Uchivectors}) can be expressed in terms of  the tetrad of the LSO  as
\begin{subequations}\begin{eqnarray}
U^a &\equiv& e_{(0)}^{\;\;\;a},\label{eq:vectors1}\\
\chi^a &\equiv& e_{(1)}^{\;\;\;a}.
\end{eqnarray}\label{eq:vectors}\end{subequations}
By inserting (\ref{eq:hab}) and (\ref{eq:vectors})  into (\ref{eq:SEMTletB}) we get the expression  for  the surface energy momentum tensor  
\begin{eqnarray}
S^{ab}= \rho e_{(0)}^{\;\;\;a}e_{(0)}^{\;\;\;b} + 
\gamma e_{(1)}^{\;\;\;a}e_{(1)}^{\;\;\;b} + \pi e_{(2)}^{\;\;\;a}e_{(2)}^{\;\;\;b},\label{eq:SEMTletf}
\end{eqnarray}
while, with $U^a$ given by (\ref{eq:vectors1}) the surface electric charge density (\ref{eq:secdSYNGE2}) may be rewritten as
\begin{eqnarray}
J^{a} &=& \psi e_{(0)}^{\;\;\;a}. \label{eq:currentSYNGEII}
\end{eqnarray} 
Now, if  we  compare the  expressions (\ref{eq:SEMTletf}) and (\ref{eq:currentSYNGEII}) (corresponding to the surface energy momentum tensor  and  the surface electric current density for a general relativistic disk) with the expression (\ref{eq:SEMTdisk}) (corresponding to the surface energy momentum tensor and the  surface electric current density for the relativistic disk-halos presented in \cite{G-PL-MQ}), we  find  that
\begin{subequations}\begin{eqnarray}
\epsilon &=& \rho= w + e,\\
\wp_1 &=& \wp_2 = \aleph\rho = p + q,\\
-\sigma&=& \psi= (1 - \aleph)^{1/2}\left( \frac{\eta^2}{p+ w} + \frac{\lambda^2}{q+e}\right)^{1/2}
\left(w + e\right)^{1/2},\label{eq:tsec}\\
\gamma &=&p+q,\\
2u^av_a&=&\frac{ \eta (2p +q)}{\lambda(p + w)} + \frac{\lambda (2q + p)}{\eta (q +e)},\label{eq:velalin}\\
\tan{2\alpha}&=& \frac{(p +w)^{1/2}(q+e)^{1/2}{2u^av_a}}{(q +e) - (p +w)}.
\end{eqnarray}\end{subequations}
Therefore, we  can conclude from the  last expression that, the rest energy $\epsilon$ as well as the pressures of the disk $\wp=\wp_1=\wp_2$, can be understood as the sum of the rest energies and pressures of the two charged perfect fluids, respectively.  The  relation (\ref{eq:velalin}) says that  the fluids  are not necessarily aligned. The  4-velocity vectors of  the  fluids, $u^a$ and $v^a$, are both time-like and pointing towards the  future, then, they satisfy the condition \cite{SYN}
\begin{eqnarray}
u^av_a=k\leq -1,
\end{eqnarray}
with $k$ an arbitrary constant.  These velocity vectors can be expressed in term of the LSO tetrad  as
\begin{subequations}\begin{eqnarray}
u^a&=&\cos(\alpha)A  e_{(0)}^{\;\;a} 
- \frac{(q+e)^{1/2}}{(p+ w)^{1/2}}\sin{\alpha} Be_{(1)}^{\;\;a},\\
v^a&=& \frac{(p+ w)^{1/2}}{(q+e)^{1/2}}\sin{\alpha}Ae_{(0)}^{\;\;a}  +
\cos{\alpha}Be_{(1)}^{\;\;a},
\end{eqnarray}\end{subequations}
with the  quantities $A$ and $B$ given by
\begin{subequations}\begin{eqnarray}
A^2=-u^{*a}u^*_a= \cos^2{\alpha} +\left(\frac{q + e}{p+w}\right)\sin^2{\alpha}
   - 2\left(\frac{q + e}{p+w}\right)^{1/2}\sin{\alpha}\cos{\alpha}u^av_a,\\
B^2= v^{*a}v^*_a= -\cos^2{\alpha} -\left(\frac{p+w}{q + e}\right)\sin^2{\alpha}
   - 2\left(\frac{p+w}{q + e}\right)^{1/2}\sin{\alpha}\cos{\alpha}u^av_a.
\end{eqnarray}\end{subequations}
The constraint
\begin{eqnarray}
2\left(\frac{p+w}{q + e}\right)^{1/2}\cos{\alpha}\sin{\alpha}|k|\geq \cos^2{\alpha} +
 \left(\frac{p+w}{q + e}\right)\sin^2{\alpha},
\end{eqnarray}
 must be  satisfied to guarantee that the vectors $u^a$ and $v^a$ are well-defined.
	
Now, for convenience, we could particularize our results for $p$, $q$, $w$ and $e$ by assuming that the pressures  and the rest energies of  the fluids are  related  by $ p=k_1q,$ and $ w=k_1e$,  $k_1$ being an arbitrary constant.
In such case we obtain
\begin{subequations}\begin{eqnarray} 
\tan{2\alpha}&=&\frac{2k_1^{1/2}|k|}{k_1 -1},\\
p&=&\aleph w,\\
q&=&\aleph e,
\end{eqnarray}\end{subequations}
and the  electric  charge  densities of the fluid satisfy the  relation
\begin{eqnarray}
{-2|k|k_1(\aleph +1)\lambda\eta} = {\aleph}{[\eta^2(2k_1 + 1) + \lambda^2(2 +k_1)k_1]}, 
\end{eqnarray}
with the  total  surface electric charge given by  the  equation (\ref{eq:tsec}).

We have presented a simple example to illustrate the method outlined in section \ref{sec:model}. In this case,  we have described explicitly the surface energy  momentum tensor  and the surface electric current of the relativistic thin disk-halos discussed in the reference \cite{G-PL-MQ} as the sum  of the surface energy momentum tensor of two charged perfect fluids with pressures $p$ and $q$ and  rest energies $w$ and $e$, respectively. In a similar way, we have also expressed the surface electric current density  of the disks as the sum of  two  surface electric current  densities with  charge densities   $\eta$ and $\lambda$  flowing with velocities $u$ and $v$ on the  surface of  the  disk. In the final part of  this  section we have particularized the calculations for a specific relation for the two pressures and between the rest energies of  the constituting fluids. With this choice, we  have obtained that the  disk can be  described as being constituted by two charged perfect fluids with a barotropic equation of  state.

\section{\label{sec:conclus}concluding remarks}
In this  work, we  derived  a  method  to  interpret the  solutions of  Einstein-Maxwell field equations  for relativistic  thin disks in terms of  two charged perfect  fluids. The  method  was  developed under the assumption that the surface energy momentum tensor as well as the electric current density of the disk can be expressed  as the sum of the energy momentum tensors and the electric current densities of two charged perfect fluids. Moreover, we have obtained explicit relations  that can be  used to express the rest energy, the  pressures and the  electric charge density on the  surface of  the disk  as a single charged fluid composed of two charged perfect fluids.  

As  an illustration of the  application of the  method, we  have described  the  solutions of the Einstein-Maxwell field  equations corresponding  to the  disk-halos discussed in the reference \cite{G-PL-MQ} in terms of two charged perfect  fluids flowing on the  surface of the disk with pressures $p$ and $q$ and rest energies $w$ and $e$, respectively. For a particular relationship between the pressures of  the  fluids,  and between the  rest energies  of  the  fluids, we  have obtained that the  disks can be  described as constituted by two charged perfect fluids with barotropic equation of  state. Here we have restricted the method to the static case. We consider that the method studied  in this work can be also used 
to  describe stationary  relativistic thin disks in terms of two charged perfect fluids. Work along this direction will soon be reported. 

\section*{Acknowledgments}
The author is thankful to C. S. L\'opez-Monsalvo, F. Nettel, A. Bravetti and H. Quevedo for  helpful remarks and hints. This  work was partially supported  by TWAS-CONACYT Postgraduate Fellowship Programme and COLCIENCIAS, Colombia.

\end{document}